\documentclass[english,aps,prb,twocolumn,groupedaddress]{revtex4-1}
\usepackage[LGR,T1]{fontenc}
\usepackage[latin9]{inputenc}
\usepackage{textcomp}
\usepackage{amssymb}
\usepackage{graphicx}
\usepackage{subscript}
\usepackage{amsmath}

\makeatletter

\DeclareFontEncoding{LGR}{}{}
\DeclareTextSymbol{\~}{LGR}{126}

\makeatother

\usepackage{babel}
\begin{document}

\title{Dynamically current-driven de Gennes - St James states in Metal-Superconductor junctions}%Differential conductance of dynamically current-driven NN'S junctions}

\author{Hervé Cercellier$^1$}
\email{herve.cercellier@neel.cnrs.fr}
\author{Hadrien Grasland$^1$}
\author{H. Q. Luo$^2$}
\author{X. Y. Lu$^2$}
\author{D. L. Gong$^2$}
\author{Z. -S. Wang$^{1,2}$}
\author{Thierry Klein$^1$}

\affiliation{$^1$Univ. Grenoble Alpes, Inst NEEL, F-38042 Grenoble, France\\
CNRS, Inst NEEL, F-38042 Grenoble, France}
\affiliation{$^2$Beijing National Laboratory for Condensed Matter Physics, Institute of Physics, Chinese Academy of Science, Beijing 100190, China}

\date{\today}
\begin{abstract}
We investigate the conductance of a Normal-Normal'-Superconductor (NN'S) junction, in which current injection destroys superconductivity in a small region N' of the superconductor, with a size varying with the applied voltage V. Voltage-dependent de Gennes - Saint James (dGSJ) bound states appearing in the N' slab lead to two distinct sets of conductance oscillations. We show that this effect significantly alters the conductance of systems for which $\kappa^2v_F \sim 10^9$ m/s such as pnictides ($\kappa$ and $v_F$ being the Ginzburg number and the Fermi velocity, respectively), and  we discuss their consequences on the identification of the bosonic modes of strongly coupled superconductors.   \end{abstract}


\maketitle

%\section{Introduction}
Current-voltage I(V) characteristics of junctions are powerful tools to study the coupling of charge carriers with the various excitations of a solid. In particular the second derivative of the differential conductance is proportional to the Eliashberg function, $\alpha^2F(\omega)$\cite{NaidyukbookPCS-Jansen1980}, that quantifies the interaction of electrons with the bosonic excitations of a given system. This quantity has hence been extensively used to study a large variety of bosonic modes, such as phonons in normal metals or magnons and crystal field splittings \cite{Khotkevich1995-Dagero2010}.  Likewise, in the strong-coupling limit, the Eliashberg theory of superconductivity predicts  an energy-dependent order parameter $\Delta(\omega)$, which exhibits strong modulations at energies for which $\alpha^2F(\omega)$ is large \cite{Carbotte1990-Scalapino1966-1965}, leading to modulations of the single-particle excitation spectrum (the so-called tunneling density of states, tDOS). One of the first and most famous examples is  the tDOS of Pb, which exhibits two dip-hump features related to the two phonon modes responsible for most of the pairing interaction \cite{Giaever1962-McMillan1965}. More recently, a well-defined satellite structure was evidenced in the tDOS of pnictides \cite{Shan2012,Wang2012}. The energy and temperature dependence of this satellite closely match the behavior of a magnetic mode observed by neutron scattering, commonly accepted to be involved in the unconventional $s^{\pm}$ gap symmetry \cite{Isonov2010-Ummarino2011-Dai2015}. 

Point-Contact Spectroscopy (PCS) is a technique in which a microscopic Normal Metal-Superconductor (NS) junction is created, for example by making a nanoscale contact between a metallic tip and the studied superconductor (the so-called spear-anvil technique which is used in our setup). At the NS interface, the superconducting pairing potential $\Delta$ is weakened by the proximity of the normal metal, which results in the formation of coherent bound states, first described by de Gennes and Saint-James \cite{dGSJ1963}, and experimentally observed in various systems \cite{Tessmer1993-Giazotto2001-Escoffier2005}. Experimentally, it turns out that in most PCS measurements the proximity effect can be neglected. However, in the case of pnictides, unusually large lifetime broadenings must be introduced to fit the conductance peaks, hindering a detailed analysis of the gap symmetry. Moreover, many PCS spectra exhibit non-negligible modulations at voltages in the range 20-60 meV which are poorly accounted for\cite{Wang2012,Daghero2011,Pecchio2013}. We show here that injection effects actually have to be taken into account in the modeling of the metal - superconductor junction in systems for which $\kappa^2v_F \sim 10^9$ m/s such as pnictides ($\kappa$ and $v_F$ being the Ginzburg number and the Fermi velocity, respectively). Indeed, in this case, a small normal region (N') develops at the NS  interface, forming a NN'S junction. The size of this N' region is dynamically driven by the applied current (and is hence voltage-dependent) significantly altering the electronic structure of the superconductor on a typical distance on the order of the superconducting coherence length. 

\begin{figure}
\includegraphics[width=8cm]{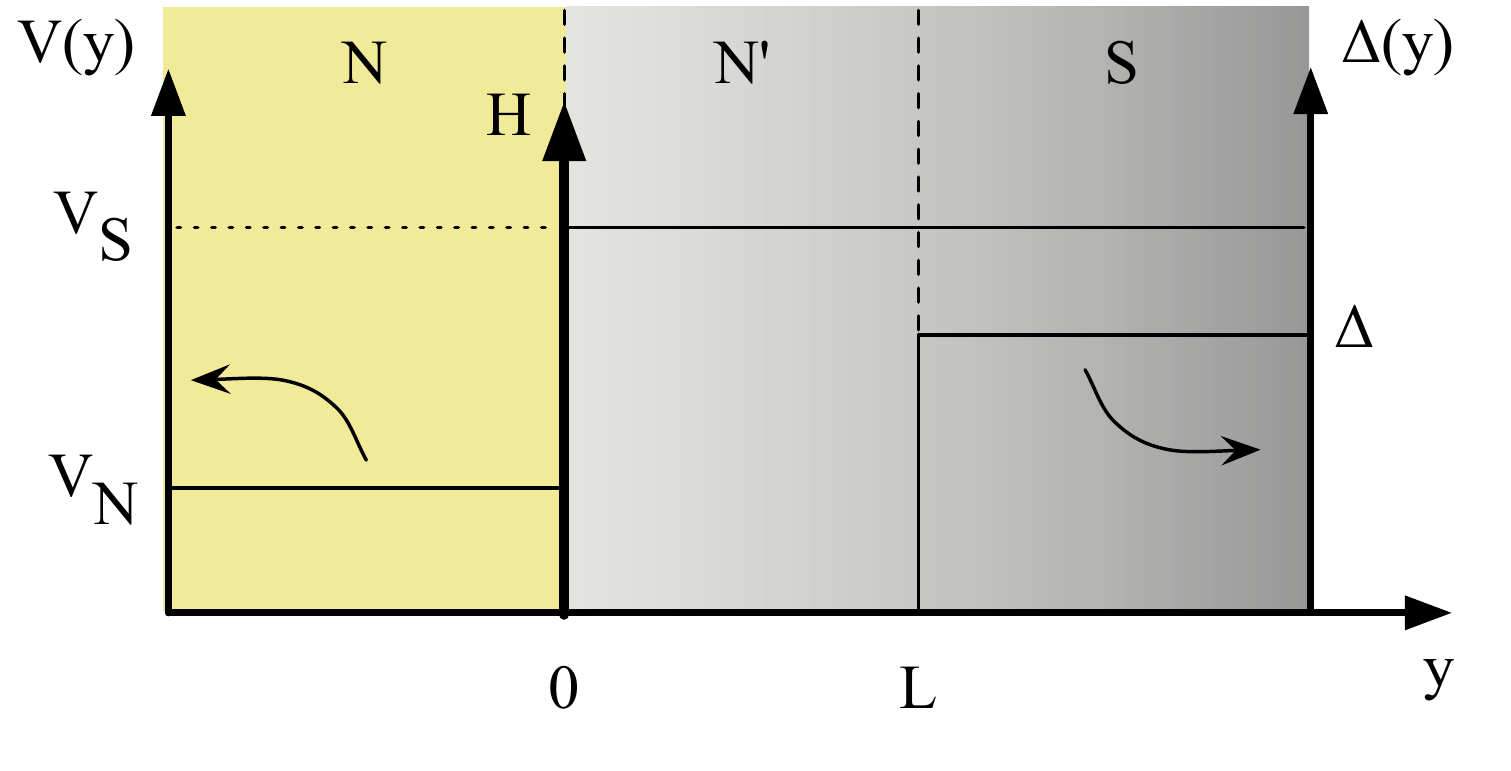}
\caption{\label{fig:schema_modele}Sketch of the NN'S junction studied in this paper. We assume that a dynamically current-driven N' normal state develops at the NS interface for $0<x<L(V)$. The superconducting pairing potential is destroyed in this region and becomes finite for $x>L$ (right scale) whereas the one-electron potential is discontinuous at the $x=0$ (left scale).}
\end{figure}

The model is based on the Bogoliubov-De Gennes equations in a simplified 1D geometry, following the approach of Hahn \cite{Hahn1985,Hahn1995}. The normal metal (superconductor) occupies the y<0 (y>0) half-space (see Fig.\ref{fig:schema_modele}). As discussed below, a metallic slab (with effective mass and Fermi velocity similar to those of the superconductor normal state) can develop in the superconductor on a size L  due to injection effects and the one-electron potential will hence be given by $V(y)=V_N+H\delta(y)+(V_S-V_N)\Theta(y)$, with $\delta(y)$ and $\Theta(y)$ the Dirac and Heaviside distributions, respectively.  Finally, the pairing potential $\Delta(y)$ can then be taken as $\Delta(y)=\Delta\Theta(y-L)$, where $\Delta$ is constant and isotropic.

The transmission coefficient of the NN'S junction will be given by (see Eq.(18) in \cite{Hahn1985}) :
\begin{eqnarray*}
T(\varepsilon,\lambda)=\frac{\tau_N(1+\tau_Nx -(1-\tau_N)x^2)}{1-2(1-\tau_N)x \cos[2(\alpha-\phi)]+(1-\tau_N)^2x^2}
 \end{eqnarray*}   
where $\tau_N=1/(1+Z_{eff}^2$) ($Z_{eff}$ being the effective transparency of the barrier), $\alpha=2\lambda\varepsilon/\pi$ with $\varepsilon=E/\Delta$ being the normalized energy in unit of the superconducting gap and $\lambda=L/\xi$ the renormalized length of N' normal slab in unit of the coherence length $\xi$,
\begin{align*}
[x, \phi] =
  \begin{cases}
    [(\varepsilon-\sqrt{\varepsilon^2-1})^2, 0]  & \text{if }  \varepsilon\geqslant 1\\
    [1,  \arccos{\varepsilon}]& \text{if }  \varepsilon<1 
  \end{cases}
\end{align*}

and the differential conductance of the junction by :
\begin{align*} 
\frac{ d I_{NS} }{ dV }= \left. \frac{ \partial I_{NS} }{ \partial V } \right|_{\lambda} + \frac{ d\lambda }{ dV } \left. \frac{ \partial I_{NS} }{ \partial \lambda } \right|_{V}
\end{align*}
where $I_{NS}(V) = - \frac{ \Delta }{ e R_N} \int_{0}^{eV/\Delta} \frac{ T(\varepsilon, \lambda) }{ \tau_N } d\varepsilon$ and the second term explicitly takes into account the dynamical change of the slab width $\lambda$. The calculation is straightforward and, introducing the renormalized voltage $\tilde{V} =eV/\Delta$, one finally obtains

%Writing :
%$$\left. \frac{ \partial I_{NS} }{ \partial V } \right|_{\lambda} = - \frac{\Delta }{ e R_N\tau_N }\frac{ \partial }{ \partial V } \int_{0}^{eV/\Delta} T(\varepsilon, \lambda) d\varepsilon  = - \frac{T(eV, \lambda)}{R_N\tau_N } $$
%and
%$$ \left. \frac{ \partial I_{NS} }{ \partial \lambda } \right|_V = - \frac{\Delta  }{ e R_N\tau_N } \int_{0}^{eV/\Delta} \frac{ dT }{ d\lambda }(\varepsilon, \lambda)~d\varepsilon$$
%one finally obtains :    
%$$G(V)=-\frac{ R_Nd I_{NS} }{dV}=\frac{T(eV,\lambda)}{\tau_N}+\frac{\Delta}{e \tau_N}\frac{d\lambda}{dV}\int_0^{eV/\Delta}\frac{dT}{d\lambda}d\varepsilon  $$
 
$$G(\tilde{V})=\frac{T[\tilde{V},\lambda(\tilde{V})]}{\tau_N}+\frac{1}{\tau_N}\frac{d\lambda}{d\tilde{V}}\int_0^{\tilde{V}}\frac{dT}{d\lambda}d\varepsilon =G_0(\tilde{V})+G_d(\tilde{V})$$
where $G_0$ corresponds to a standard "static" conductance channel, which is non-zero in all possible geometries, and $G_d$ a "dynamical" channel which exists only when the slab width changes with voltage. The differential conductance of the junction is then fully determined by the Blonder-Tinkham-Klapwijk (BTK) barrier\cite{BTK1982} ($\tau_N$) and the voltage dependence of the slab width, $\lambda(\tilde{V})$. The fixed $L$ geometry ($G(V)=T(eV,L)/\tau_N$)  has been thoroughly studied in the literature and we will rather focus on the case for which the slab width L changes with applied voltage due to high current injection that will drive the junction out of equilibrium (see discussion below).  Note that the former equations have been derived at T=0 and, at finite temperature, the differential conductance is given by the standard convolution by the Fermi function.The temperature hence acts as a low-pass filter on the differential conductance, strongly smearing out oscillations with periods smaller than a few $k_B T$.

\begin{figure}
\includegraphics[width=9cm]{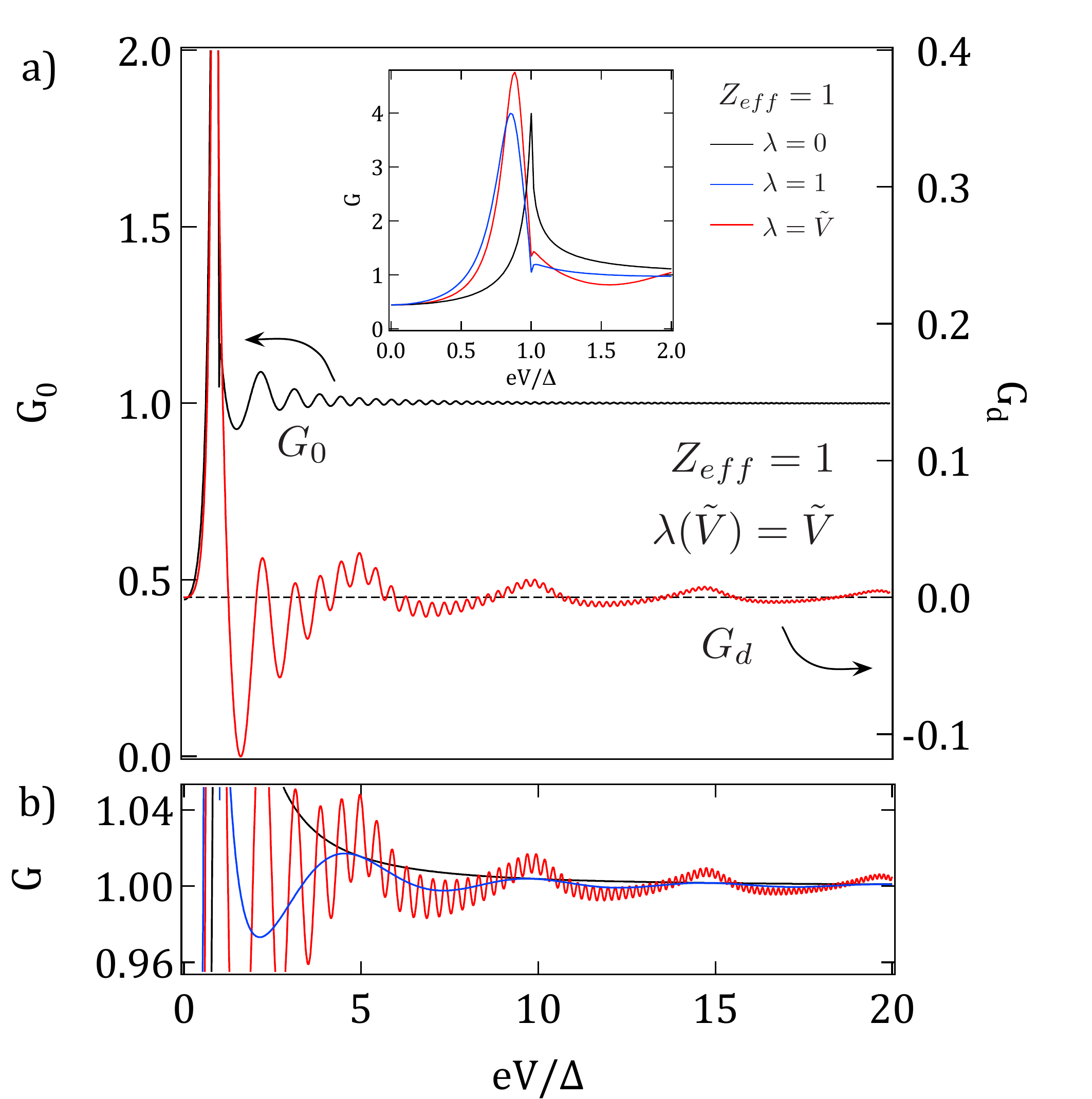}
\caption{\label{fig:G0&Gd} (a) $G_0$ and $G_d$ components of the normalized differential conductance in a dynamically current-driven NN'S junctions (see text for details) calculated for $Z_{eff}=1$ and a slab growth law $\lambda(\tilde{V})=1\times \tilde{V}$. Both components exhibit a high-frequency oscillation, whereas a low-frequency oscillation is also present in the dynamical channel $G_d$  (see text). Inset : comparison of the low biais spectra  in the standard BTK model (no N' slab, black curve),  a static N' slab of size L=$\xi$ (blue curve) and a dynamical N' slab of width $L(V)=eV\xi/\Delta$ (Red Curve). Note that the coherence peak is shifted towards low biais voltages in the presence of the slab. (b) high biais zoom the total normalized differential conductance for three same geometries.}
\end{figure}

 Fig.\ref{fig:G0&Gd}a displays the $G_0$ and $G_d$ components of the differential conductance at T=0, calculated for a moderate barrier $Z_{eff}=1$ and a linearly growing slab of width $L=\alpha eV$ with a typical $\alpha$ value on the order of  $\xi/\Delta$ i.e. $\lambda=\beta\tilde{V}$ with $\beta=1$. For this $Z_{eff}$ value, $G_0$ exhibits a low-bias peak and an oscillating behavior for $\tilde{V}>1$ with a period and amplitude that decrease with increasing voltage. As expected, $G_0$ tends towards 1 at high biases. Similarly, $G_d$ also exhibits a low-bias peak and oscillations which coincide with the $G_0$ ones but, the high-bias $G_d$ baseline is centered around zero conductance, as it is mainly a correction to the excess current caused by the dynamical change of the N' slab. Besides, a second, low-frequency oscillation is visible, with a slowly decreasing amplitude.   
 
In the standard BTK model (i.e. no N' slab), the conductance is featureless except for the quasiparticle peak located exactly at the gap energy ($\tilde{V}=1$, see inset of Fig.\ref{fig:G0&Gd}a), but it is important to note this peak is shifted to lower biases in the presence of the N' slab (either of constant width or for a dynamically increasing slab). Indeed, the De Gennes-Saint James (dGSJ) subgap states present in the N' slab \cite{dGSJ1963} give rise to oscillations in the differential conductance (cosine term in the transmission coefficient) and, in presence of the slab, $G$ is maximum when $\alpha - \phi =0$ [$\pi$]. A first subgap state is hence obtained for $\tilde{V}=cos(2\lambda \tilde{V} / \pi)\leq 1$ and the corresponding broad conductance peak is a signature of the discontinuity in the dGSJ density of states and not of a true quasiparticule peak anymore (see inset of Fig.2a). The peak position still gives directly the energy position of the dGSJ feature for constant $L$,  but does no longer correspond to any particular energy in the case of a dynamically growing N' slab. It is hence important to note that the differential conductance can no longer be directly related to the density of states of the junction as the electronic structure of the slab dynamically evolves with the applied voltage.

For $\tilde{V}>1$, the conductance maxima occur for voltages $\tilde{V}=n \pi^2/2 \lambda(\tilde{V})$, which correspond to the well-known Rowell-McMillan (RMM) oscillations \cite{Rowell1966}. Those oscillations are periodic for a constant $L$ value (see Fig.\ref{fig:G0&Gd}b) and are present in both $G_0$ and $G_d$ for the dynamically current-driven slab with a quasi-period that decreases as $\lambda$ increases. The low-frequency oscillations visible in $G_d$ have a more subtle origin : they correspond to  the entrance of a new dGSJ bound state below the superconducting gap and are defined by : $\lambda(\tilde{V})=n \pi^2/2$. As the slab size increases, the total number of subgap states inside the slab increases, with sudden jumps appearing when a new state enters. The differential conductance of a dynamically growing NN'S junction will hence show two kinds of characteristic features, both related to the existence and the evolution of dGSJ bound states inside the N' region (see Fig.2b). First,the quantized nature of the dGSJ spectrum will give rise to broad subgap states and non-periodic RMM type oscillations above the gap. Second, the modulation of the total number of bound states inside the N' region as it grows with voltage will lead to a low-frequency, slowly decaying oscillation.

\begin{figure}
\includegraphics[width=9cm]{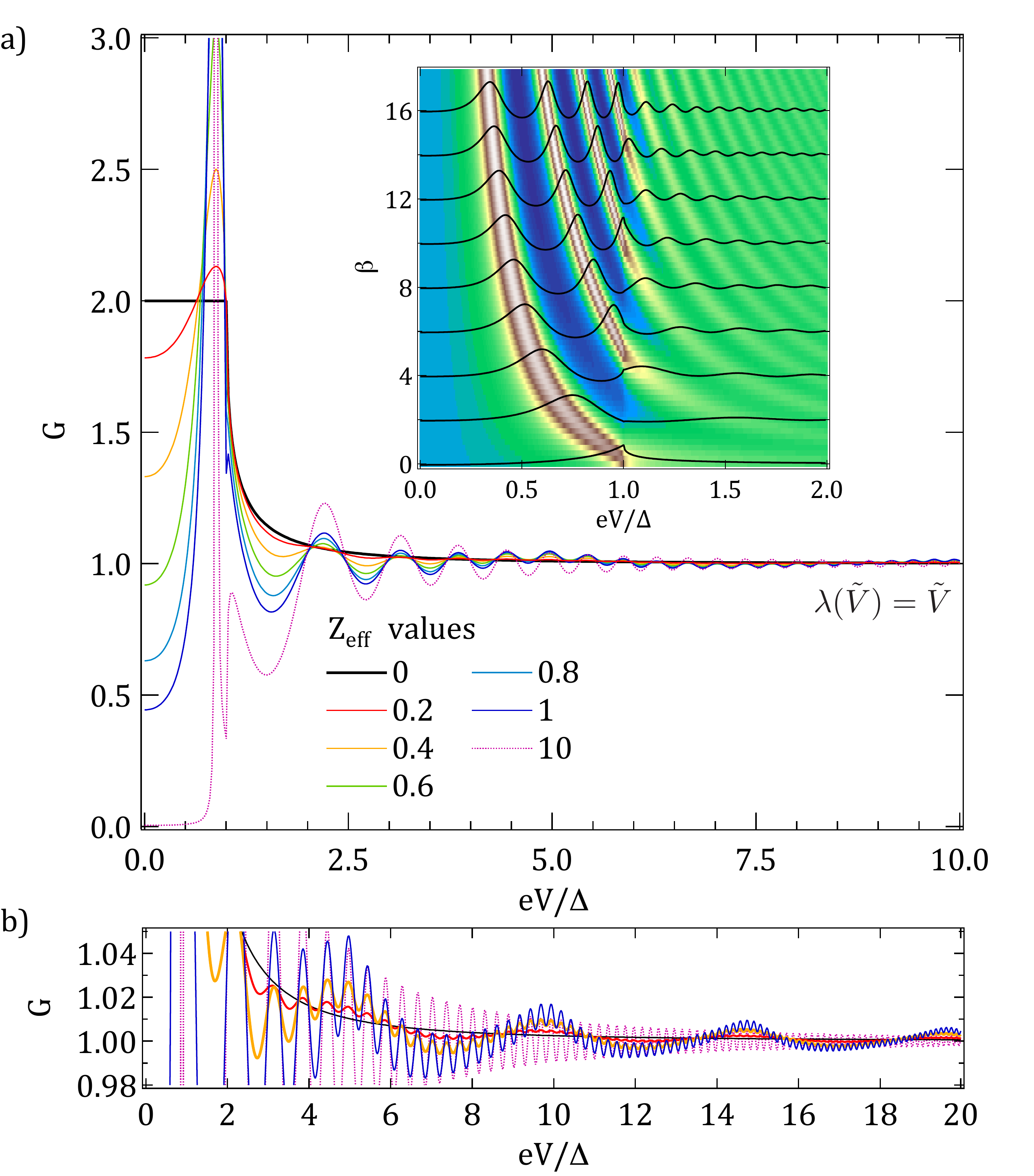}
\caption{\label{fig:effectofZ}(a)Effect of the effective barrier parameter $Z_{eff}$ on the differential conductance of a dynamically current-driven NN'S junctions with $\lambda = 1\times \tilde{V}$. Inset : influence of the N' region growth rate $\beta=d\lambda/d\tilde{V}$ on the differential conductance at T=0. The spectra corresponding to the values displayed on the left axis are displayed as full black lines above the color map. (b) High bias zoom of the differential conductance showing that the amplitude of the low frequency oscillations increases for $Z_{eff}\rightarrow 1$.}
\end{figure}

The effect of the interface effective barrier $Z_{eff}$ on the differential conductance of the dynamical NN'S model is displayed in Fig.\ref{fig:effectofZ}. As expected from the BTK model, the zero-bias conductance decreases with increasing $Z_{eff}$, as the probability of Andreev reflexion at the NN' interface drops, and the shape of the conductance becomes more and more tunnel-like. The oscillations vanish for $Z_{eff}=0$, as the differential conductance tends towards the "bare" BTK prediction. Indeed, in this case, the charge carriers can cross the NN' interface without being scattered or reflected, and no quasiparticle interference can occur : the NN'S sandwich behaves as an effective NS junction. For increasing $Z_{eff}$ quasi-particule interferences become more and more effective, resulting in sharper subgap peaks and RMM oscillations of higher amplitude. As shown in Fig.3b, the low frequency oscillations increases for $Z_{eff}\lesssim1$ but are hidden by the RMM signal for higher barrier strengths. For $Z_{eff}=10$ the low-frequency signal is completely invisible compared to the $1/\varepsilon^{2}$ envelope function of the RMM oscillations.

Finally, the influence of the grow rate $\beta=d\lambda/d\tilde{V}$ is displayed in the inset of Fig.\ref{fig:effectofZ}a. As discussed above, for $\beta \leq 4$ the main low biais conductance peak shifts towards lower energies with increasing $\beta$ and do not correspond to the quasiparticle peak anymore. For higher $\beta$ values, several dGSJ subbands appear before the gap voltage is reached and the differential conductance exhibits n subgap peaks. Correspondingly, the quasi period of the high bias oscillations decreases with increasing $\beta$.

 \begin{figure}[h]
\includegraphics[width=8.5cm]{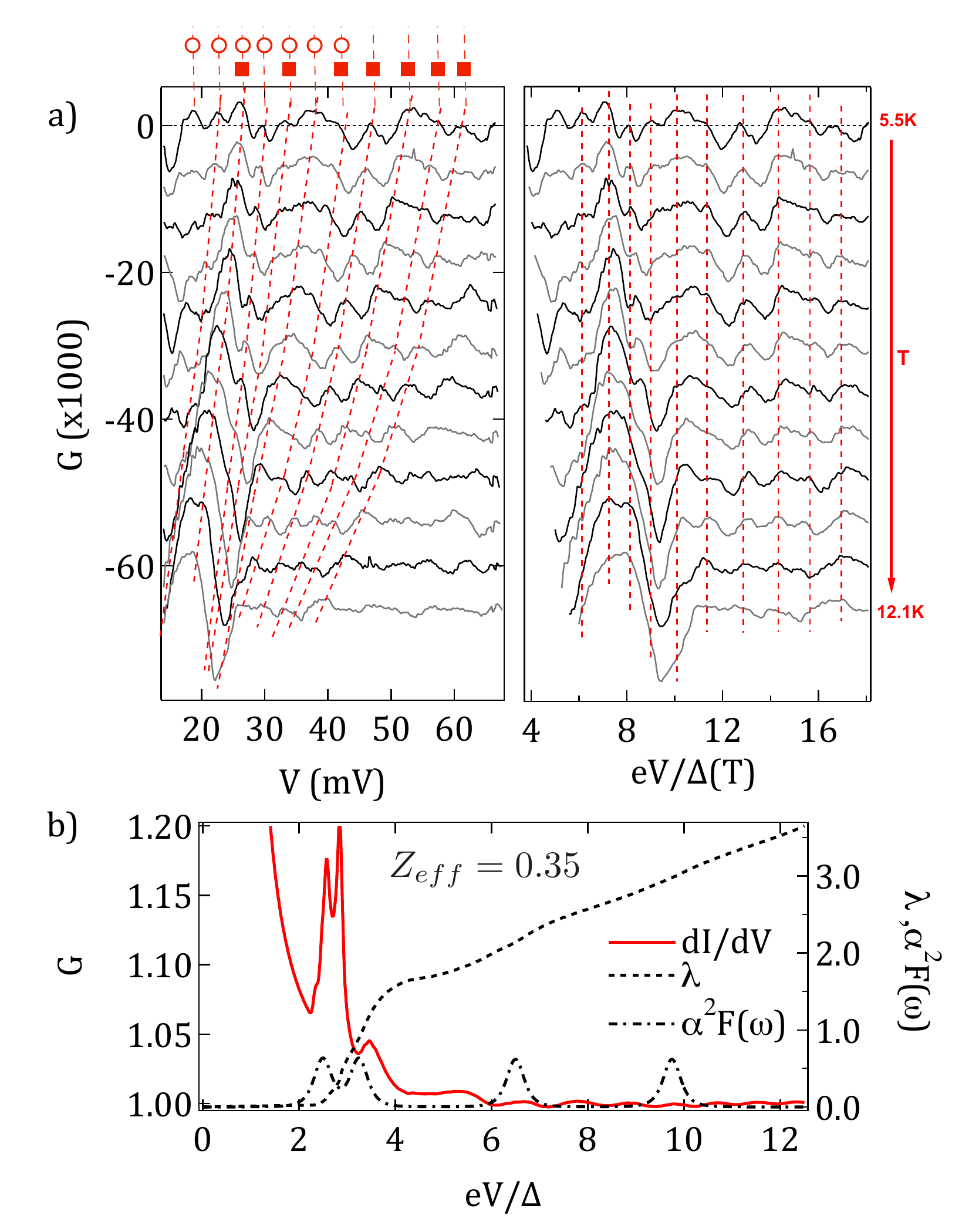}
\caption{\label{fig:Mesure}  (a) High bias differential conductivity measurements in a BaFe$_{2-x}$Ni$_{x}$As$_2$ close to optimal doping between $T=5.5$ K and $12.1$ K (the different curves have been shifted for clarity). As shown quasi-periodic strong modulations are visible (left) and the position of the maxima roughly scales as the superconducting gap (right). The position of the peaks can be indexed following two sequences scaling as $V_{max}=\alpha n^{2/3}$ (see text for details) with $\alpha \sim 10$ meV (open circles) and $\alpha \sim 16$ meV (closed squares), respectively. (b) Calculated $\lambda(\tilde{V})$ (dotted line) for typical parameters of the 122 pnictides (see text). The red full line is the corresponding calculated differential conductance at T=0. Strong modulations of the differential conductivity are visible in the vicinity of the bosonic modes.}
\end{figure}

Let us now discuss the two main mechanisms that can destroy superconductivity in the vicinity of the NN' contact . The model will be derived for the case of a point contact between the N and N' regions, and a hemispherical N' region (which we call a "normal bubble"). It turns out, that the dGSJ states for a s-wave superconductor in spherical symmetry have the same characteristic energies and lengths as the 1D case \cite{Gunsenheimer1996}. Besides, taking into account the 2D or 3D nature of the junction often leads to results that can be nearly reproduced by a 1D model with an effective barrier strength \cite{Khotkevich1995-Dagero2010}. Apart from some amplitude deviations, we then expect the 1D model to be a fair approximation of a more realistic 3D point-contact.     

First, if the local current density exceeds the depairing current density $J_d$ in a region of typical size $L\approx \xi$, it becomes energetically favorable for the sample to transit into the metallic state in this region. Let us consider a planar NS junction of contact resistance $R_N$ and assume that the current is injected in the S electrode through a point contact and spreads isotropically in the superconductor from this point source (this is a good approximation for small $Z_{eff}$). In this approximation the surfaces of constant current densities are hemispheres  centered around the injection point and $J\sim J_d$ for $L\sim \sqrt{V/[2\pi R_NJ_d]}$. Note that we assumed here a linear $I(V)$ characteristics which leads to an error of at most a factor 2 in the $Z_{eff} \rightarrow 0$ limit (for $\tilde{V}<1$), and becomes exact in the $\tilde{V}\gg1$ limit.

The RMM maxima of the differential conductance will then be obtained for $\tilde{V}=(n^2\pi^5\xi^2R_NJ_de/2\Delta)^{1/3}$. Taking $J_d(0) \sim 10^{-10}/\lambda_L(0)^2\xi(0)$ and $\Delta(0)=\hbar v_F/\pi\xi(0)$ one obtains :
$$ \tilde{V}\sim [n^210^8R_N/v_F\kappa^2]^{1/3}F(T/T_c)$$
with $\kappa =\lambda_L/\xi$ is the Ginzburg parameter and $v_F$ being the Fermi velocity. Note that  $F(T/T_c) \sim 1$ so that $\tilde{V}$ is only very weakly temperature dependent. Similarly, the low frequency oscillations will be observed for $\lambda=n\pi^2/2$ i.e. for $\tilde{V}= [n^210^8R_N/v_F\kappa^2]$. Taking $R_N \sim 10$ $\Omega$, those out-of-equilibrium effects are hence expected o be observed in systems for which $v_F\kappa^2\sim 10^9$ m/s. Note that in this normal bubble scenario the energy period of the oscillations \textit{shrinks} with temperature (following the gap) in strong contrast with the $\Delta(\omega)$ related bosonic spectrum in superconducting tunneling experiments, which \textit{shifts} in energy with temperature, as the superconducting gap closes. 
 
In standard superconductors (such as Nb) $\kappa \sim 10$, $v_F \sim 3.10^5$ m/s and $10^8R_N/v_F\kappa^2 \sim 30$; the oscillations are hence barely visible but in 122-pnictides, such as BaFe$_{2-x}$(Co,Ni)$_{x}$As$_2$, $\kappa \sim 100$ (see for instance \cite{Rodiere2012}), $v_F \sim 10^5$ m/s\cite{Brouet2009} and $10^8R_N/v_F\kappa^2 \sim 1$ making this system very sensitive to those current induced out of equilibrium phenomena. This system has been previously intensively studied in our group (see [17] and references therein for sample details) and we have performed  PCS measurements (with $R_N \sim 10-20$ $\Omega$) in BaFe$_{2-x}$Ni$_{x}$As$_2$ single crystals close to optimal doping. As shown in Fig.\ref{fig:Mesure}(a), as expected, clear oscillations are visible in the high bias differential conductivity in very good agreement with our theoretical current-driven N' "bubble" (the low energy spectra present the typical Andreev bump usually observed in those systems, not shown).  Those oscillations have been observed in several samples and in all of them the conductance maxima could be ascribed to two quasi periods $V_{max} = \alpha n^{2/3}$ with $\alpha \sim 10$ meV and $\alpha \sim 16$ meV, respectively (see Fig.4a). Introducing the two gaps of this system ($\Delta_S\sim 4$ meV and $\Delta_L\sim 10$ meV \cite{Wang2012,Pecchio2013}), one hence gets $[10^8R_N/v_F\kappa^2] \sim 0.1$ in very reasonable agreement with the electronic parameters of this system \cite{Brouet2009}. Note that, as discussed above, the presence of those oscillations also leads to a shift (and important smearing) of the low biais peak which then leads for an underestimation of the gap value on the order of $\sim 10\%$ for the small gap and up to $\sim 30\%$ for large gap. This shift/smearing effect can account for the large dispersion of the gap values deduced from PCS measurements \cite{Pecchio2013}. Finally note that, as expected, the position of the observed maxima shrinks with T, nearly following the temperature dependence of the gap (see right panel in Fig.4a).

The second mechanism which is expected to give rise to an injection-dependent normal bubble is the formation of a non-thermal "hot spot" near the point-contact. In this scenario, multiple scattering of high-energy injected electrons leads to an out-of-equilibrium electron distribution function that does not allow for a superconducting gap to exist on a typical scale that depends on injection current and electron-boson coupling. In order to calculate the $\lambda(\tilde{V})$ law, numerical calculations were carried out using the formalism introduced by Hahn \cite{Hahn2002}. The input parameters of the model are the Fermi velocity, the coherence length (or superconducting gap), the density of states at the Fermi energy  and the electron-boson coupling Eliashberg function. The $\lambda(\tilde{V})$ law calculated for  typical parameters of the 122 Ni or Co doped pnictides \cite{Brouet2009} ($N_0=1.3$ $10^{20}$$J^{-1}.nm^{-3}$,  $\xi_0=3 nm$, $\Delta_0=4 meV$) and the Eliashberg fonction sketched by the dashed-dotted line is displayed in Fig.\ref{fig:Mesure}b (dotted line) together with the corresponding calculated differential conductance (red line). 

As shown, If $\alpha^2F(\omega)$ exhibits well-defined strongly coupled low-energy peaks $\lambda(\tilde{V})$, the injected electrons are poorly scattered for low voltages and the normal bubble remains very small. When the injection voltage reaches the first (sharp) bosonic mode, electron scattering becomes very efficient and the normal bubble grows quickly. As the voltage increases, more and more bosonic modes will become available for scattering, resulting in a smoothly growing bubble, with some modulations for voltages matching strongly coupled bosonic modes. This non-trivial bubble growth is then expected to give rise to satellite structures in the differential conductance (see red curve in fig.  \ref{fig:Mesure}b) opening the possibility to identify the relevant bosonic modes for charge transport even in imperfect, strongly diffusive junctions. Even though most of the single-electron spectroscopic information is lost in such junctions, this work suggests that in real experiments electron-boson coupling can still be probed accurately in "bad" junctions, provided the quasiparticle mean free path is of the order of magnitude of, or slightly larger than the superconducting coherence length. Unfortunately, in our case, the strong oscillations induced by the first mechanism are hindering the clear identification of those bosonic modes but it is worth noting the enhanced oscillations observed for $V \sim 20$ meV which are consistent with the magnetic modes deduced from neutron measurements \cite{Isonov2010-Ummarino2011-Dai2015} (see also \cite{Wang2012}).
       
We have presented here a model for the differential conductance of current-driven NN'S junctions, in which the N' slab is dynamically induced by current injection effects. This slab can be due either to the fact that the injected current density exceeds the depairing current in the vicinity of the injection point or to downscattering of high-energy charge carriers by strongly coupled bosonic modes. Critical current effects are expected to play an important role in systems for which  $v_F\kappa^2\sim 10^9$ m/s in very good agreement with PCS spectra performed in BaFe$_{2-x}$Ni$_{x}$As$_2$ signal crystals. If the electron mean free path is comparable to the Pippard coherence length, enhanced modulations of the differential conductivity are expected to be observed for voltages corresponding to strongly coupled bosonic modes.

This research was performed using funding from the Lab Alliance for
Nanosciences and Energies for the Future (Lanef), Institut Néel of
the Centre National de la Recherche Scientifique, and Université Grenoble
Alpes. The work at IOP, CAS, is supported by NSFC (No.11374011) and CAS (SPRP-B: XDB07020300).
This work is supported by the French National Research Agency through Grant No. ANR-12-JS04-0003-01 SUBRISSYME.
\bibliographystyle{apsrev4-1}
%\bibliography{\string"/media/hadrien/ECHANGE/Biblio/Jabref/BDD Jabref\string"}

\end{document}